\documentclass[twocolumn,10pt]{asme2e}

\usepackage{graphicx}
\usepackage{xcolor}
\usepackage{hyperref}
\hypersetup{
    colorlinks=true,
    linkcolor=blue,
    citecolor=blue,
    urlcolor=blue,
}

\confshortname{ICEF 2021}
\conffullname{the ASME 2021 \\ The Internal Combustion Engine Fall Conference}

\confdate{October 13 – 15}
\confyear{2021}
\confcity{Virtual Conference, Online}
\confcountry{USA}

\papernum{ICEF2021/69657}

\title{Machine Learning Approaches to Learn HyChem Models}
\author{Weiqi Ji
    \affiliation{
	Department of Mechanical Engineering\\
	Massachusetts Institute of Technology\\
	Cambridge, MA 02139\\
    Email: weiqiji@mit.edu
    }
}

\author{Julian Zanders                             
    \affiliation{
	Department of Electrical Engineering and Computer Science\\
	Massachusetts Institute of Technology\\
	Cambridge, MA 02139\\
    Email: jzanders@mit.edu
    }
}

\author{Ji-Woong Park
    \affiliation{Argonne National Laboratory\\
	Lemont, IL 60439\\
    Email: ji.park@anl.gov
    }
}

\author{Sili Deng \thanks{Address all correspondence to this author.}
    \affiliation{Department of Mechanical Engineering\\
	Massachusetts Institute of Technology\\
	Cambridge, MA 02139\\
    Email: silideng@mit.edu
    }
}

\begin{document}

\maketitle    

% TODO: change Tables to real Tables (optional).

%%%%%%%%%%%%%%%%%%%%%%%%%%%%%%%%%%%%%%%%%%%%%%%%%%%%%%%%%%%%%%%%%%%%%%
\begin{abstract}
{\it The HyChem (Hybrid Chemistry) approach has recently been proposed for modeling high-temperature combustion of real, multi-component fuels. The approach combines lumped reaction steps for fuel thermal and oxidative pyrolysis with detailed chemistry for the oxidation of the resulting pyrolysis products. Determining the pyrolysis submodel requires extensive experimentation on speciation measurements. Recent work has been directed to learn HyChem from an existing HyChem model for a similar fuel, which requires less data. However, the approach usually shows substantial discrepancies with experimental data within the Negative Temperature Coefficient (NTC) regime, as the low-temperature chemistry is more fuel-specific than high-temperature chemistry. This paper proposes a machine learning approach to learn the HyChem models that can cover both high-temperature and low-temperature regimes. Specifically, we develop a HyChem model using the experimental datasets of ignition delay times covering a wide range of temperatures and equivalence ratios. The chemical kinetic model is treated as a neural network model, and we then employ stochastic gradient descent (SGD), a technique that was developed for deep learning, for the training. We demonstrate the approach in learning the HyChem model for F-24, which is a Jet-A derived fuel, and compare the results with previous work employing genetic algorithms. The results show that the SGD approach can achieve comparable model performance with genetic algorithms but the computational cost is reduced by 1000 times. In addition, with regularization in SGD, the SGD approach changes the kinetic parameters from their original values much less than genetic algorithm and is thus more likely to retrain mechanistic meanings. Finally, our approach is built upon open-source packages and can be applied to the development and optimization of chemical kinetic models for internal combustion engine simulations.

% One of the key steps in developing HyChem models is determining the stoichiometric coefficients and rate constants of the fuel pyrolysis steps. Those parameters are usually determined by sequentially matching the experimental data in shock tubes and flow reactors. While these divide-and-conquer approaches have achieved success in the kinetic modeling of many fuels, the determination of the parameters requires intuition and expertise. This work presents a machine learning approach to determine the HyChem model parameters which can simultaneously optimize the HyChem model using all relevant corrupted datasets, including speciations in shock tubes, flow reactions as well as ignition delay times. We incorporate HyChem models into neural ordinary differential equations and predict the speciation data by solving the neural ordinary differential equations. We then leverage our recently developed differential combustion simulation package Arrhenius.jl powered by Julia to backpropagate the error between predictions and experimental data to HyChem model parameters. The HyChem model parameters are then optimized with stochastic gradient descent, which has been successfully demonstrated in optimizing deep learning models. We demonstrate the approach in learning HyChem models for diesel fuels. This work shall open the possibilities of differential programming in learning and optimizing complex combustion kinetic models and benefit the development of compact and accurate kinetic models for internal combustion engine simulations.
}
\end{abstract}

%%%%%%%%%%%%%%%%%%%%%%%%%%%%%%%%%%%%%%%%%%%%%%%%%%%%%%%%%%%%%%%%%%%%%%
% \begin{nomenclature}
% \entry{A}{You may include nomenclature here.}
% \entry{$\alpha$}{There are two arguments for each entry of the nomemclature environment, the symbol and the definition.}
% \end{nomenclature}

%%%%%%%%%%%%%%%%%%%%%%%%%%%%%%%%%%%%%%%%%%%%%%%%%%%%%%%%%%%%%%%%%%%%%%
\section*{INTRODUCTION}

Predictive chemical kinetic models for real fuels play an important role in modeling the combustion processes in practical engines. Modeling real fuels is very challenging since real fuels are comprised of a variety of components with different physical and chemical properties. Therefore, much effort has been directed to developing compact models for real fuels that are computationally efficient and can accommodate the variations in fuel compositions. These approaches can be categorized as the "surrogate modeling approach" and the "HyChem approach."

In the surrogate model approach \cite{violi2002experimental, eddings2005formulation, dooley2010jet}, it is assumed that the functional groups characterize the combustion properties of a fuel mixture. More specifically, if a real fuel and a surrogate mixture share similar compositions of functional groups, they should also share similar combustion behaviors. Besides the functional groups, the physical properties of the mixture, such as heat capacity and molecular weights, also play important roles. Previous studies \cite{violi2002experimental, eddings2005formulation, dooley2010jet} have demonstrated that using a handful of pure hydrocarbon fuels could formulate a surrogate fuel mixture that effectively reproduces the combustion behaviors of a complex real fuel, including the ignition delay times (IDTs) and laminar flame speeds. However, it is often an art to determine the choice of surrogate fuel components and the procedures for developing surrogate fuels.

Recently, Wang and his co-workers have proposed an end-to-end modeling approach named "HyChem (Hybrid Chemistry)" \cite{wang2018physics, xu2018physics, xu2020physics}. The HyChem approach describes the fuel pyrolysis and the subsequent oxidation of the pyrolysis products in two submodels. The fuel pyrolysis/oxidative pyrolysis is modeled by several experimentally constrained and lumped reaction steps. The oxidation of the pyrolysis products is modeled by a detailed foundational chemistry model. The two submodels are interconnected. The pyrolysis process provides "reactants" for the oxidation process; the oxidation provides heat and radicals to facilitate the endothermic pyrolysis of the fuel.

% The stoichiometric coefficients and reaction rate parameters of these lumped reactions are experimentally constrained, are mainly determined via a sequential optimization using species profiles. While this method is systematic and provide accurate fuel modeling, it involves extensive experimental measurements. Therefore, recent efforts \cite{ryu2021data} have been directed to construct the lumped reaction steps for a new fuel based on the fuel with similar physical/chemical property through optimization method. 

The lumped reaction steps are mainly determined via a sequential optimization using species profiles during the high-temperature pyrolysis. Compared to the species profiles that require expensive laser diagnostic equipment, ignition delay times are easier to measures in shock tubes and rapid compression machines. Therefore, recent work \cite{ryu2021data} has proposed learning the rate constants of lumped reaction model from an ignition delay times dataset only by employing the stoichiometric coefficients from existing models for similar fuels. However, the base model usually shows large discrepancies in the low-temperature regimes, as the low-temperature chemistry is more fuel-specific compared to high-temperature chemistry. Thus, one has to optimize the lumped reaction steps from experimental data such that the optimized model can cover both high- and low-temperature regimes. Ryu et al. \cite{ryu2021data} employed genetic algorithms and IDT data to develop a data-driven kinetic model for the Jet fuel of F-24 using the Jet-A HyChem model as the base model. The core of such a data-driven modeling approach is the optimization step using genetic algorithms. However, genetic algorithms (GA) usually suffer from the curse of dimensionality, and thus they are usually limited to less than one hundred parameters. In addition, it is usually questionable how well a data-driven model is able to extrapolate beyond the training dataset.

Therefore, this work aims to tackle the challenges in achieving scalability and generalization of the data-driven HyChem model using techniques developed for training deep neural network models. We propose the stochastic gradient descent (SGD) optimizer to learn the pyrolysis submodel in HyChem models. The SGD optimizer is widely applied in the deep learning communities for its efficiency as well as scalability in dealing with large datasets, and its robustness in optimizing high-dimensional non-convex neural network models. Our recent work \cite{ji2021autonomous} has also shown that a chemical reaction network is equivalent to a neural network with a single hidden layer. Similarly, solving ordinary differential equations (ODEs) of reaction network models is equivalent to solving infinite-depth deep residual networks \cite{chen2018neural}, which further rationalizes exploiting SGD in training HyChem models. However, the base model is constrained using the sequential optimization from multiple datasets, and thus changing the kinetic parameters too much has the risk of over-fitting the experimental dataset. Therefore, we also adopt various regularization techniques used in training deep neural network models to increase the generalization performance of learned models. Thus, we can achieve a model that has comparable performance to the conventional GA approach, but with much smaller changes in the kinetic parameters from the reference values.

However, one of the major obstacles for exploiting SGD in optimizing combustion models is the lack of software ecosystems that can efficiently and accurately compute the gradient of simulation output to model parameters. For instance, the finite difference method (often termed the "brute-force method") usually suffers both computational inefficiency due to the cost scaling with the number of parameters and inaccuracy due to the truncation error \cite{baydin2018automatic}. Conversely, gradient evaluation methods based on auto-differentiation (AD) have shown both efficiency and accuracy in the training of large-scale deep neural network models. Many open-source AD packages have been developed in the last decade, including TensorFlow \cite{abadi2016tensorflow} and Jax \cite{jax2018github} backed by Google, PyTorch \cite{paszke2019pytorch} backed by Facebook, ForwardDiff.jl \cite{revels2016forward} and Zygote.jl \cite{innes2018don} in Julia. To this end, this work presents a simulation package Arrhenius.jl and techniques that enable SGD for combustion model optimization, and applies them to the modeling of F-24, which is a variant of Jet-A with additives.

%%%%%%%%%%%%%%%%%%%%%%%%%%%%%%%%%%%%%%%%%%%%%%%%%%%%%%%%%%%%%%%%%%%%%%
\section*{METHODS}

%%%%%%%%%%%%%%%%%%%%%%%%%%%%%%%%%%%%%%%%%%%%%%%%%%%%%%%%%%%%%%%%%%%%%%
\subsection*{Arrhenius.jl}

\begin{figure*}
    \centering
    \includegraphics[width=0.8\textwidth]{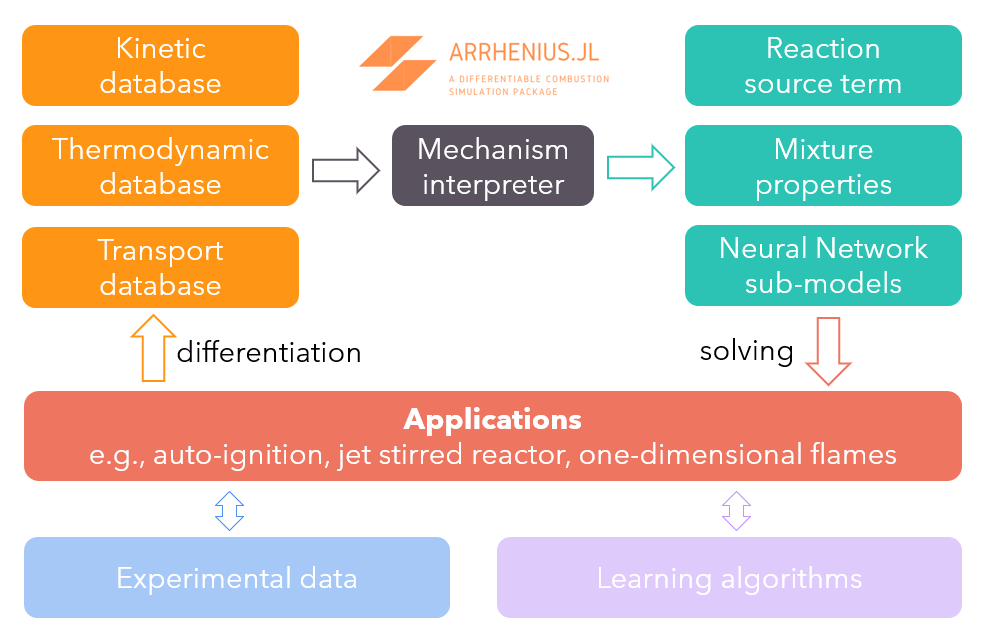}
    \caption{Schematic diagram showing the structure of the Arrhenius.jl \cite{jiarrheniusgithub} package.}
    \label{fig:schem}
\end{figure*}

Arrhenius.jl is built using the programming language of Julia to leverage the rich ecosystems of auto-differentiation and differential equation solvers. Arrhenius.jl does two types of differentiable programming: (i) it can differentiate elemental computational blocks. For example, it can differentiate the reaction source term with respect to kinetic and thermodynamic parameters as well as species concentrations; (ii) it can differentiate the entire simulator in various ways, such as solving the continuous sensitivity equations \cite{ji2019evolution} as did in CHEMKIN \cite{kee1989chemkin} and Cantera \cite{goodwin2009cantera} and in adjoint methods \cite{rackauckas2018comparison, rackauckas2017differentialequations}. The first type of differentiation is usually the basis of the second type of high-level differentiation. Arrhenius.jl offers the core functionality of combustion simulations in native Julia programming, such that users can readily build the applications on top of Arrhenius.jl and exploit various approaches to do high-level differentiations.

Figure \ref{fig:schem} shows a schematic diagram of the structure of the Arrhenius.jl package. The Arrhenius.jl reads in the chemical mechanism files in YAML format maintained by the Cantera community, and the chemical mechanism files contain the kinetic models, thermodynamic, and transport databases. The core functionality of Arrhenius.jl is to compute the reaction source terms and mixture properties, such as heat capacities, enthalpies, entropies, Gibbs free energies, etc. In addition, Arrhenius.jl offers flexible interfaces for users to define neural network models as submodels and augment them with existing physical models. For example, one can use a neural network submodel to represent the unknown reaction pathways and exploit various scientific machine learning methods to train the neural network models, such as neural ordinary differential equations \cite{chen2018neural,rackauckas2020universal} and physics-informed neural network models \cite{raissi2019physics,ji2020stiff}. One can then implement the governing equations for different applications with these core functionalities and solve the governing equations using classical numerical methods or neural-network-based solvers, such as physics-informed neural networks \cite{raissi2019physics, ji2020stiff}. Arrhenius.jl provides solvers for canonical combustion problems, such as simulating the auto-ignition in constant volume/pressure reactors and oxidation in jet stirred reactors.

Compared to the legacy combustion simulation packages, Arrhenius.jl can not only provide predictions given the physical models, but also optimize model parameters given experimental measurements. By efficiently and accurately evaluating the gradient of the solution outputs to the model parameters, experimental data can be incorporated into the simulation pipeline to enable data-driven modeling with deep learning algorithms.

\subsection*{Dataset and Base Model}

Following the work of \cite{ryu2021data}, we employ the IDT dataset compiled in \cite{ryu2021data}, including the IDTs of the lean, stoichiometric, and rich F-24/air mixtures measured at 20 bars in a rapid compression machine (RCM) and a shock tube (ST) over the temperature range of ~625 K to ~1250 K. The RCM is driven pneumatically with a hydraulic stop apparatus, and the initial chamber temperature and pressure are achieved by adjusting the compression ratio \cite{allen2012application}. Low-temperature (up to 700 K) IDTs were measured in the RCM. The shock tube uses a polycarbonate diaphragm to separate the driver and driven sections, the lengths of which are 2.744 m and 4.227 m, respectively. A tailored mixture of He and N$_2$ was used as the driver gas, and premixed F-24/air mixtures were introduced to the driven section.

For the base model, we adopt a HyChem version of the Jet-A skeletal mechanism \cite{xu2018physics} that covers the NTC regime. It has 48 species and 254 reactions. In addition, details of the formulations and stoichiometric coefficients in the F-24 model are the same as in \cite{ryu2021data}, and they are listed in Table \ref{tbl:pathway}. For simplicity and consistency, all of the simulations are carried out under constant volume conditions using Arrhenius.jl.

\begin{table*}
  \caption{List of reaction pathways in the pyrolysis submodel presented in the skeletal Jet-A HyChem model \cite{xu2018physics}}
  \label{tbl:pathway}
  \begin{center}
      \includegraphics[width=0.5\linewidth]{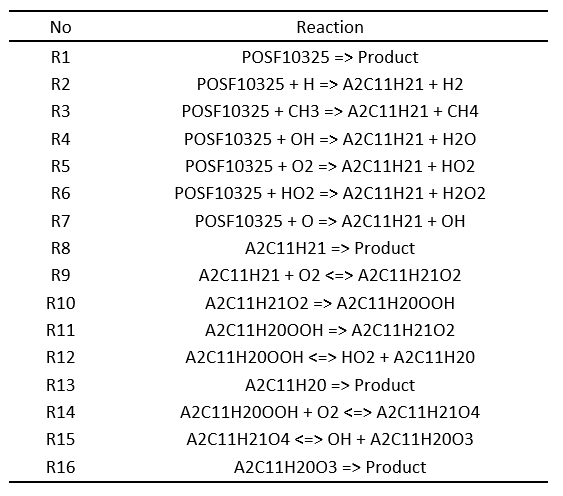}
  \end{center}
\end{table*}

\subsection*{Gradient Evaluation}

As the only available dataset is the IDT, the key step in utilizing SGD for the optimization of the HyChem model is to compute the gradient of the IDT with respect to the kinetic parameters. Recent work \cite{gururajan2019direct, ji2019evolution, lemke2019adjoint, almohammadi2021tangent} have substantially advance the algorithms for computing the gradient of IDT to model parameters, this work employs the sensBVP method \cite{gururajan2019direct}. The approach converts the initial value problem (IVP) to a boundary value problem (BVP) by treating the temperature at ignition as a boundary condition and the IDT as a free variable to solve. The algorithm is implemented with Arrhenius.jl and further enhanced with the auto-differentiation, multi-threading, and sparse matrix techniques. For the size of the skeletal HyChem model considered in this work, the computational cost of the sensitivity of IDT is negligible comparing to computing the IDT.

It is worthy to note that the gradient of laminar flame speed to kinetic parameters is also usually computed using the boundary value problem approach as in the sensBVP approach for IDTs. Thus, the computational cost for the gradient computation of flame speed is also negligible comparing to computing the flame speed. Therefore, the gradient-based optimization approach proposed in this work can be readily extended to the experimental dataset of flame speed as well.

\subsection*{Optimization Techniques}

Following the HyChem approach, the 16 steps of the pyrolysis submodel, listed in Table \ref{tbl:pathway}, are subject to optimization, and the rest of the core model is fixed. Although in the work of \cite{ryu2021data}, only the pre-factor $A$ and activation energy $Ea$ are optimized, the SGD optimizer can easily scale up to high-dimension, such that all three parameters, including the non-exponential temperature dependence factor $b$, are optimized. Consequently, the optimization involves 48 parameters.

As discussed in the introduction, we would like to regularize the model to achieve an optimized model that can fit the experimental datasets well but only moderately change the kinetic parameters from the base model. \textcolor{black}{To facilitate regularization, the model parameters are scaled and centered at zero, such that the relative changes of Arrhenius parameters is to be optimized. Specifically, we define the scaled parameters as
\begin{equation}
    p = [ln(A/A_0), b-b_0, Ea - Ea_0],
    \label{eq:pdef}
\end{equation}
where the subscript $_0$ refers to the base model. The dimension of $Ea$ is specified as $kcal/mol$ rather than $cal/mol$ as a change of 2 $kcal/mol$ in $Ea$ is roughly equivalent to changes of $e$ times in $A$, such that the changes in $A$ and $Ea$ will be balanced and avoid stiffness in the parameter space.}

The loss function is defined as the mean square error (MSE) between the predicted IDTs in the logarithmic scale using \textcolor{black}{the optimized model and the measured IDTs, as shown in Eq.~\ref{eq:loss}:
\begin{equation}
    Loss = MSE\left(log(IDT^{op}), log(IDT^{exp}) \right)
\label{eq:loss}
\end{equation}}
The choice of logarithmic scale is similar to the choice of relative error between measured and predicted IDT as in \cite{ryu2021data}. The Adam \cite{kingma2014adam} optimizer with the default learning rate of 0.001 is adopted. Weight decaying and early stopping \cite{bengio2017deep} are employed to regularize the parameters, such that the optimization will prefer kinetic parameters that are close to their original values.

%%%%%%%%%%%%%%%%%%%%%%%%%%%%%%%%%%%%%%%%%%%%%%%%%%%%%%%%%%%%%%%%%%%%%%
\section*{RESULTS}

In this section, we first present an overview of the learning results and then present the details of the learning algorithms. The code implementations can be found in the Github repository for Arrhenius.jl \cite{jiarrheniusgithub}.

\subsection*{Overview of Learning Results}

\begin{figure}
\begin{center}
\includegraphics[width=1.0\linewidth]{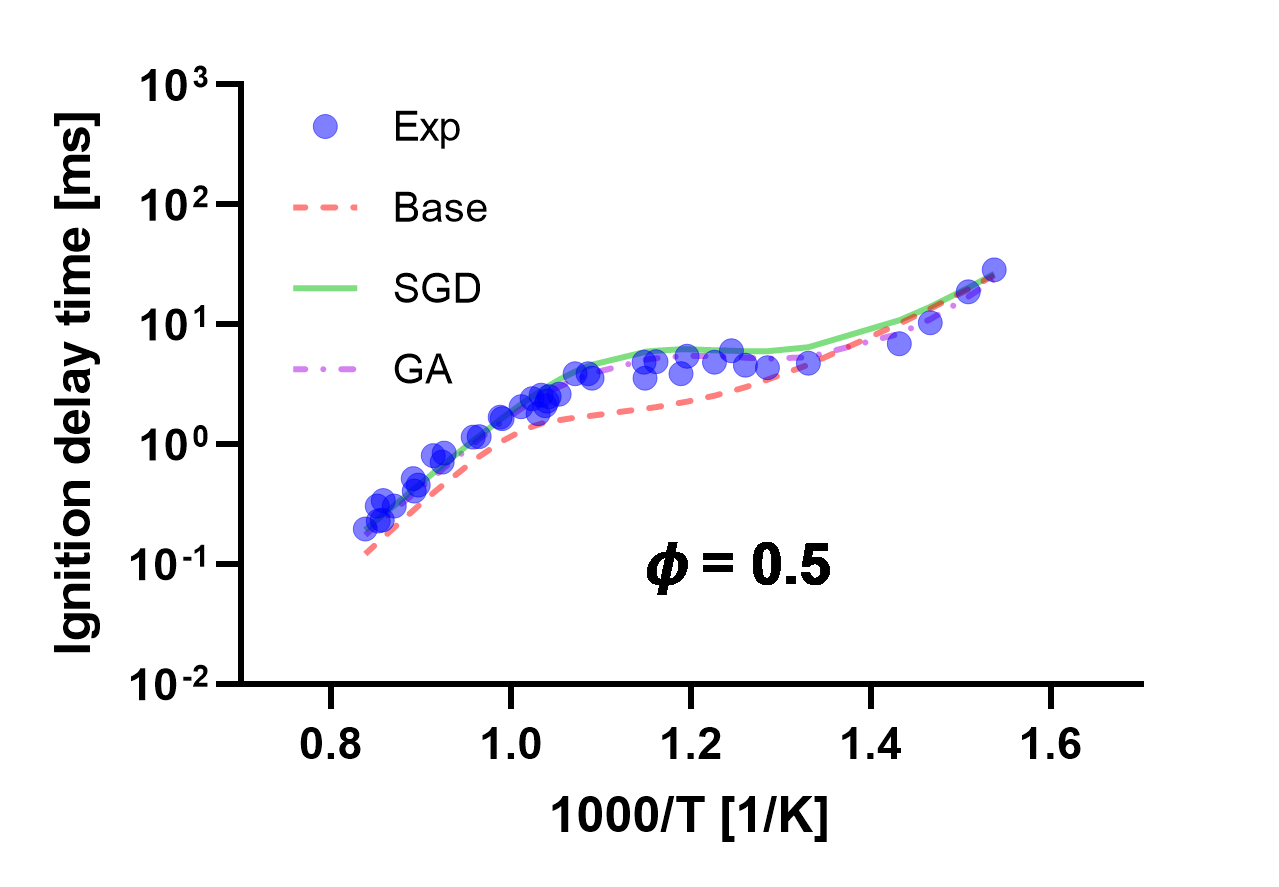}
\includegraphics[width=1.0\linewidth]{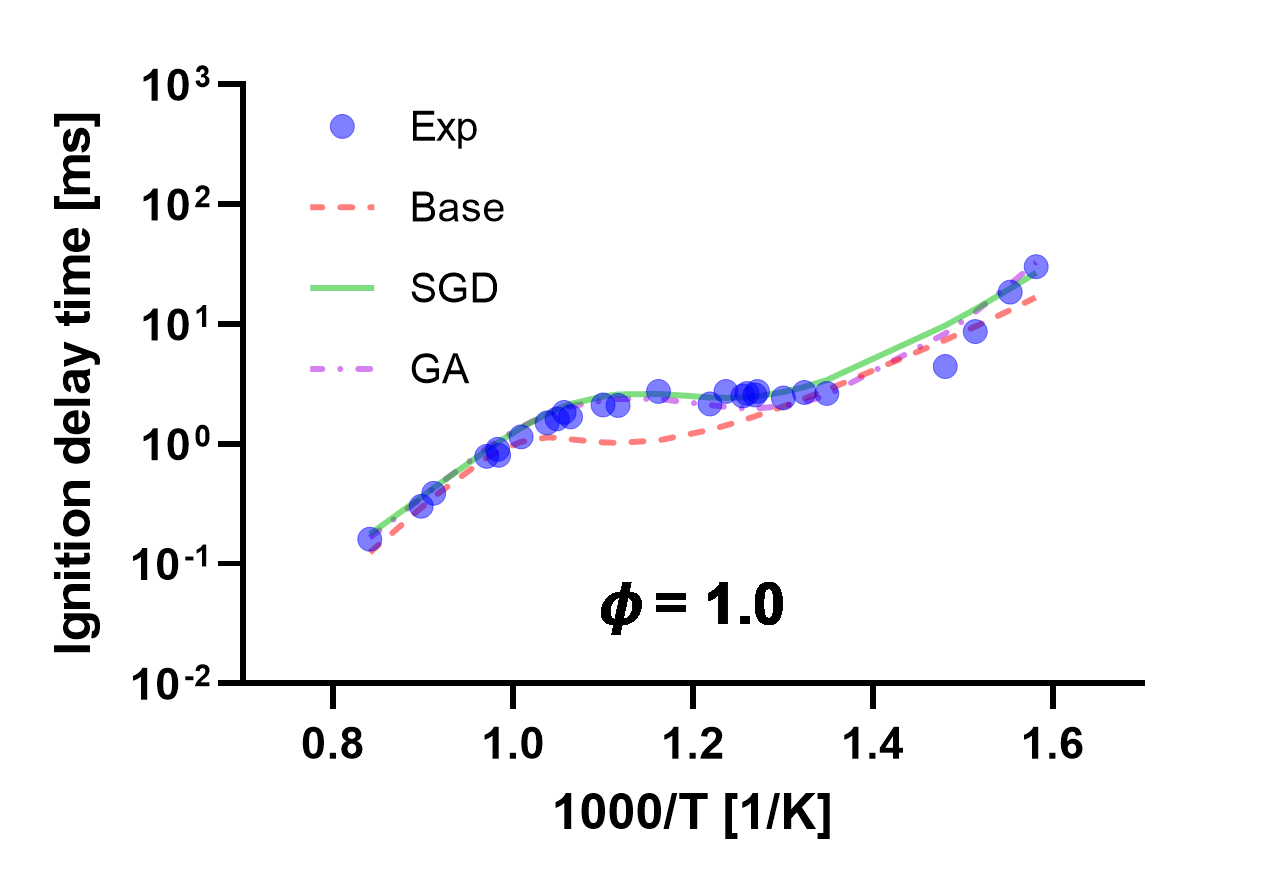}
\includegraphics[width=1.0\linewidth]{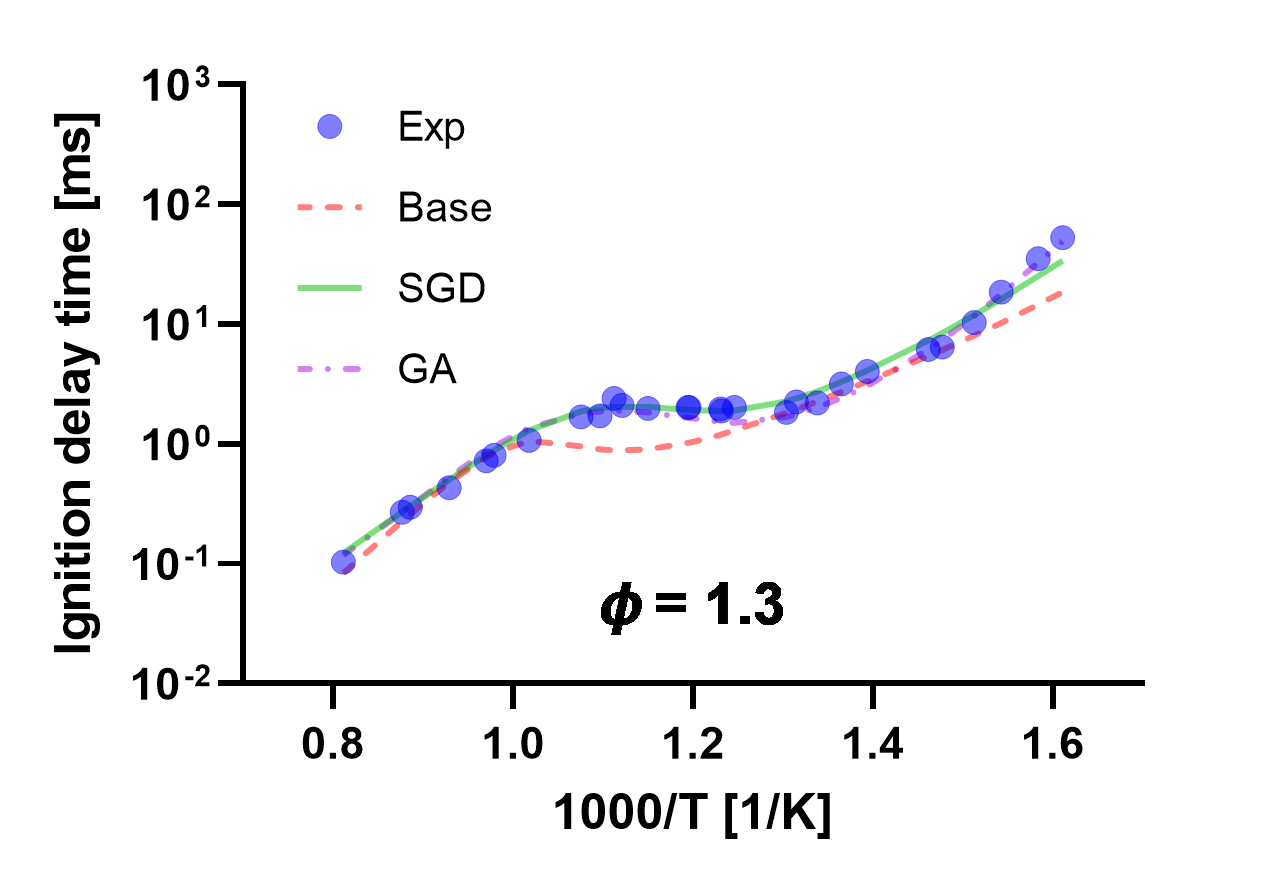}
\end{center}
\caption{Ignition delay times from F-24 RCM/shock tube measurements (Exp), and predicted by the skeletal Jet-A HyChem model (Base), the optimized HyChem model using SGD in current work (SGD), and optimized HyChem model using genetic algorithms in \cite{ryu2021data} (GA), at 20 bars for lean ($\phi = 0.5$), stoichiometric, and rich ($\phi = 1.3$) fuel/air mixtures.}
\label{figure:IDT}
\end{figure}

Figure \ref{figure:IDT} show the comparisons between the experimentally measured IDTs and the predictions using the base HyChem model, the optimized model using genetic algorithms (GA), and the current model. As pointed out by \cite{ryu2021data}, the base model substantially underpredicts the IDTs in the NTC regimes for all three equivalence ratios, while it performs reasonably well in the high-temperature regimes. Conversely, both of the optimized models substantially improve the performance in the NTC regime.

Since the slope of the response curve is indicative of the global activation energy, it is interesting to note that the current model shows a smaller global activation energy at lower temperature regimes, where the temperature is below 700 K, than the one suggested by the experimental data and the GA model. Such global activation energy is related to the key reactions that govern the low-temperature auto-ignition \cite{ji2017crossover, zhao2013role}, and thus it is important to accurately predict the global activation energy to gain physical insight of the ignition behavior of F-24. The discrepancies in the predicted global activation energy can be attributed to several plausible reasons:

\begin{enumerate}
    \item The current model indeed underpredicts the activation energy due to large uncertainties in the training data, since the IDT measurements in the NTC regime are usually associated with large uncertainties. In addition, the simulations for training and predictions are carried out under constant volume assumptions for simplicity and consistency, which might not be appropriate for two-stage ignition \cite{zhang2016first}. These uncertainties in the training dataset and modeling may result in a biased model. In such case, the data re-balance procedure in \cite{ryu2021data} can be employed to increase the weights of low-temperature data on the model training.
    
    \item The reported measurements are subject to heat loss, and therefore overestimate the global activation energy. The heat loss for experiments with long testing time in RCM usually has a substantial effect on the IDTs. The reported experimental data using the temperature at the end of compression gives higher global activation energy than the actual one. In such case, the effective temperature \cite{he2005experimental,ji2015intermediate} is usually a better representative temperature than the temperature at the end of compression.
\end{enumerate}
On the other hand, the SGD optimization tends not to over-fit the data at low temperatures compared to the GA optimization, since multiple regularization techniques are employed to prevent over-fitting. From this point of view, the experimental data are likely giving an overly large global activation energy than the actual one. Nevertheless, good data-driven modeling not only can help us develop a predictive model from experimental data but also help reveal potential experimental uncertainties and guide further experimental designs to advance our physical understandings.

\begin{figure}
\begin{center}
\includegraphics[width=1.0\linewidth]{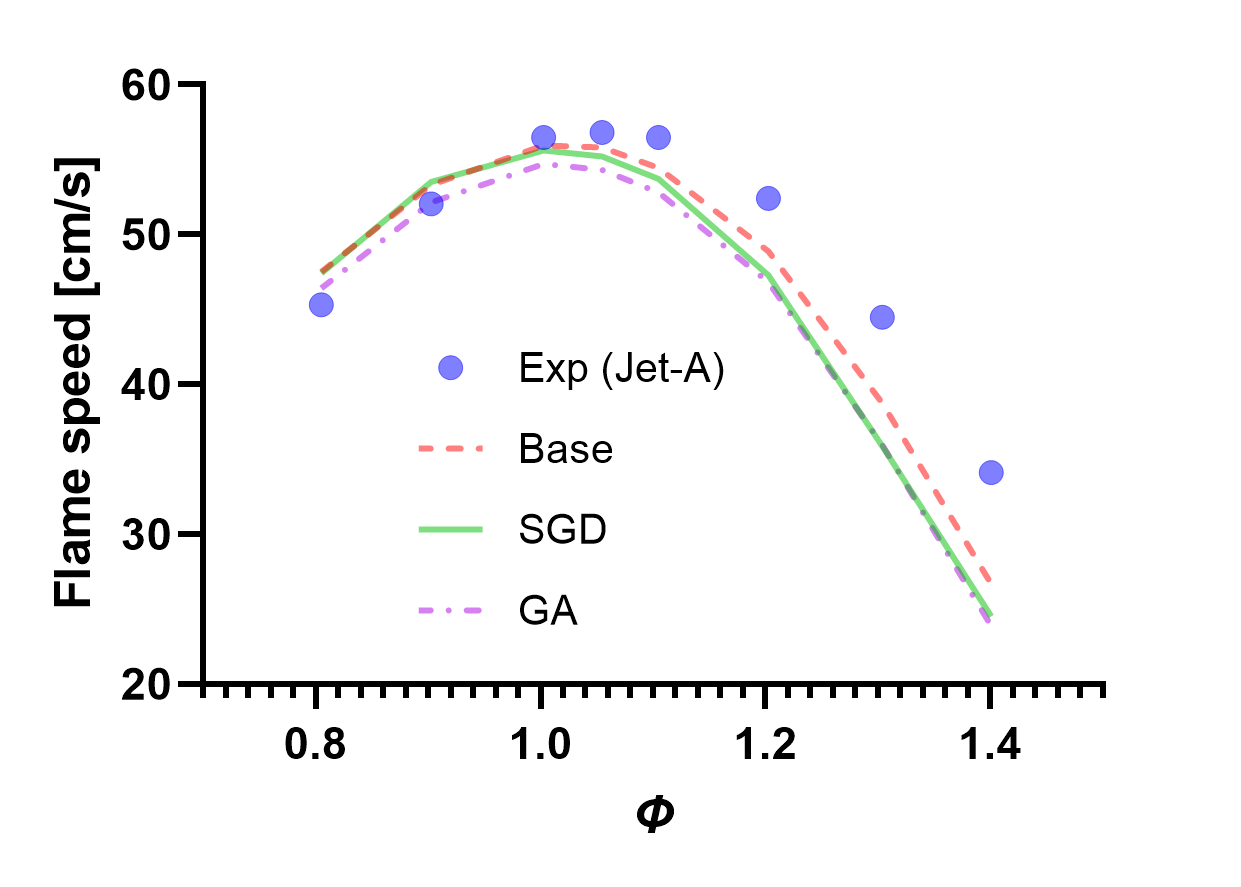}
\end{center}
\caption{Laminar flame speeds from Jet-A measurements \cite{xu2018physics} (Exp), and predicted by the skeletal Jet-A HyChem model (Base), the optimized HyChem model using SGD in current work (SGD), and optimized HyChem model using genetic algorithms in \cite{ryu2021data} (GA), at one atm and 423 K.}
\label{figure:SL}
\end{figure}

To assess the generalization capability of the learn model, we further evaluate the performance of the optimized models in predicting laminar flame speeds, although the model is not optimized against flame speed data. To the best of our knowledge, there are no experimental data of flame speed for F-24 in the open literature. Thus, we apply the model to predict flame speed for Jet-A as F-24 is derived from Jet-A and might share similar flame propagation behaviors. The predicted and measured flame speeds for Jet-A are presented in Fig. \ref{figure:SL}.  

While it is difficult to assess which model performs better given the uncertainties in the experimental data, we can use the base model as a reference, for the base model is calibrated for high-temperature pyrolysis. The base model is expected to perform well for flame propagation that is dominated by high-temperature chemistry. Overall, both SGD and GA models predict the similar flame speed with the base model, with the differences within 2 cm/s. On the other hand, the GA model systematically predicts slower laminar flame speeds than the base model. The predictions by the current model agree with those by the base model at fuel-lean conditions and show similar trends as those by the GA model under fuel-rich conditions. This could be potentially attributed to the regularization in the SGD optimization. As shown in Tables \ref{tbl:arrheniusparam} and \ref{tbl:p}, the kinetic parameters in the current model are closer to the base model compared to those in the GA model. For the GA model, the activation energy for the direct decomposition reaction of R1 is changed a lot, and such changes may lead to systematic changes in flame speed under both fuel-lean and rich conditions. For the SGD model, it is possible that the flame speed is only sensitive to the changes of kinetic parameters under fuel-rich conditions, as the fuel-lean and rich conditions correspond to the different radical pools. Future work could exploit advanced pathway analysis tools, such as global pathway analysis \cite{gao2019global}, to further investigate the different trends under lean and rich conditions. Nevertheless, the SGD model tends to less affect the model parameters for high-temperature chemistry while improving the low-temperature chemistry from experimental data. Thus, comparing to the GA model, the SGD model is more likely to be able to extrapolate to conditions beyond the range of the training dataset.

While the experimental dataset of laminar flame speed can be readily incorporated in the SGD approach as discussed in the methodology section, the results in Fig. \ref{figure:SL} suggest that learning the HyChem model from the ignition delay time dataset might be sufficient to develop a model that works well for both ignition and flame propagation with the help of regularization. Such generalization capability is especially useful recognizing that experimental measurements of laminar flame speed are usually limited to near-atmospheric pressures due to the onset of turbulence at elevated pressures \cite{long2019numerical, ji2020dependence}. Nevertheless, the SGD approach can potentially enable us to develop a data-driven kinetic model by providing only the high-pressure ignition delay time dataset but can work well for predicting both ignition and flame propagation at engine relevant elevated pressures.

\begin{table*}
  \caption{The kinetic parameters of the pyrolysis submodels in the skeletal Jet-A HyChem model \cite{xu2018physics} (Base), the optimized model using SGD in current work (SGD), and the optimized model using genetic algorithm in \cite{ryu2021data} (GA).}
  \label{tbl:arrheniusparam}
  \begin{center}
      \includegraphics[width=0.7\linewidth]{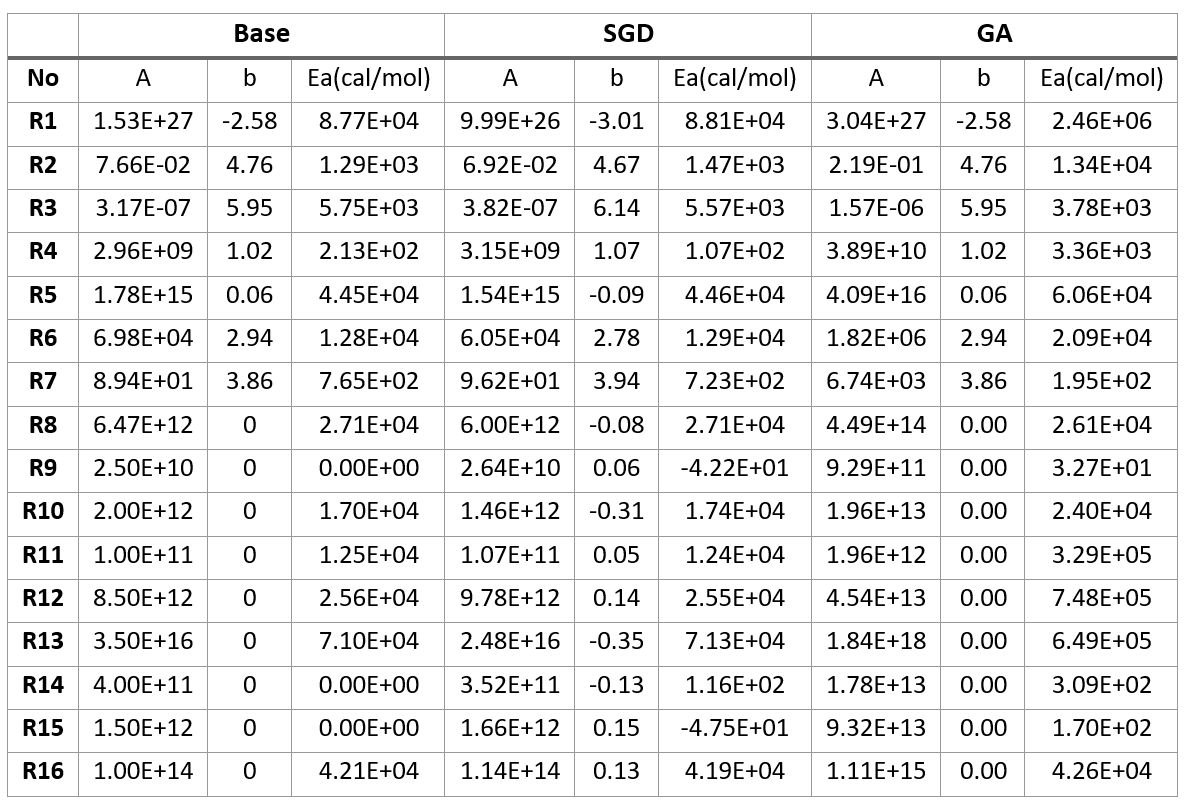}
  \end{center}
\end{table*}

\begin{table*}
  \caption{The changes of the kinetic parameters in the optimized models in current work (SGD) and in \cite{ryu2021data} (GA) with respect to the base model of a skeletal Jet-A HyChem model \cite{xu2018physics}.}
  \label{tbl:p}
  \begin{center}
      \includegraphics[width=0.7\linewidth]{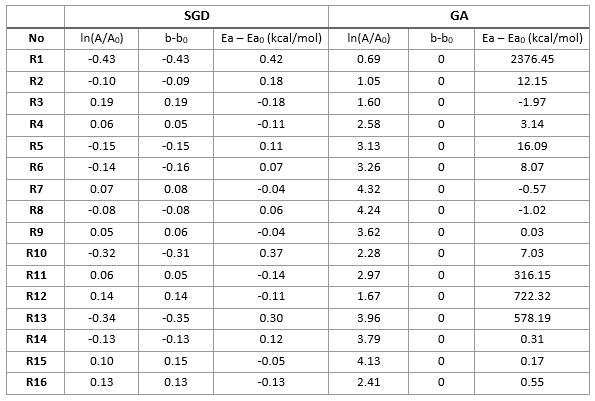}
  \end{center}
\end{table*}

\subsection*{Details of Learning Algorithms}

\begin{figure}
\begin{center}
\includegraphics[width=1.0\linewidth]{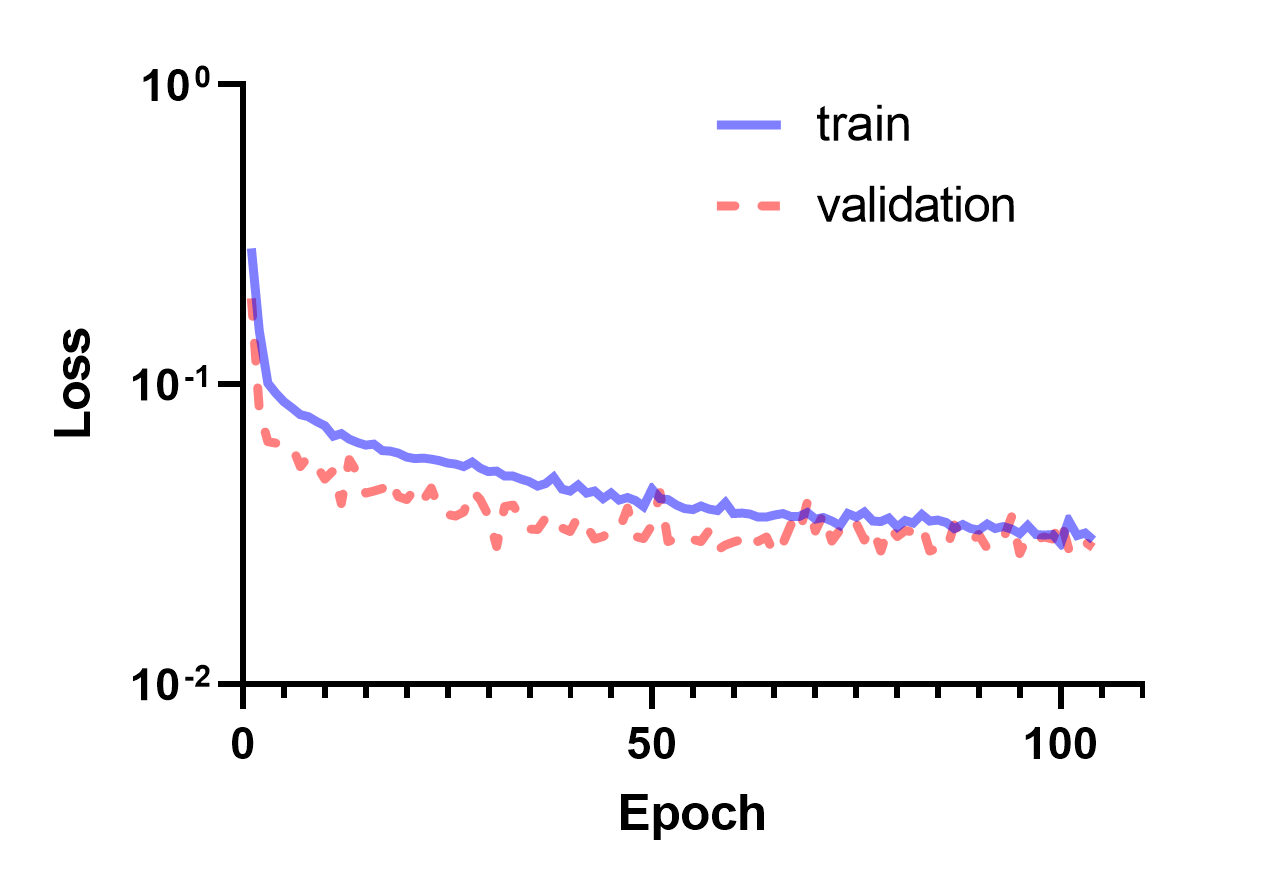}
\end{center}
\caption{The history of the loss function in the training (with 76 samples) and validation (with 20 samples). Within each epoch, the training dataset was randomly shuffled and the optimizer updated the model parameters for each sample and 76 times for each epoch.}
\label{figure:loss}
\end{figure}

We then present the details of the learning algorithms that produce the SGD-optimized model. Figure \ref{figure:loss} shows the loss history in the training and validation of the model. 76 out of 96 samples were randomly chosen as the training dataset, and the rest were for the validation. Within each epoch, the training dataset was randomly shuffled to introduce randomness to the training process to prevent over-fitting. Mini-batch optimization was adopted such that the model parameters were updated for each sample and 76 times for each epoch. Cross-validation and early-stopping were exploited to prevent over-fitting. For instance, the training is stopped when the loss in the validation reaches a plateau, although the loss in the training is still decreasing. The validation loss reached a plateau at around 60 epochs, although we let the training continues to 110 epochs to confirm the plateau. We thus select the model at 58 epochs, which is the closest checkpoint near 60 epochs, and the checkpoint is saved every time the validation loss reaches a new minimal. In summary, we adopted multiple techniques, including mini-batching, cross-validation and early-stopping, to regularize the model and prevent over-fitting. The regularization in SGD in general leads to better generalization performance, which is one of the most important advantages comparing to GA optimization, although both SGD and GA can give a similar fitness in a small dataset with less than one hundred parameters.

% \textcolor{red}{The $L_2$ norm of the model parameters also reaches a plateau at around 60 epochs, which suggests a local minimal with a stiff landscape is reached at around 60 epochs. While the choice of optimal model in SGD optimization is still an art, a recent study suggests that a stiff landscape usually indicates a good generalization performance \cite{fort2019stiffness}. [Weiqi: I am not confident about this part. If this part is hard to understand, we can delete the figure for p and those discussions ...]}

% \begin{figure}
% \begin{center}
% \includegraphics[width=1.0\linewidth]{./figs/pnorm.png}
% \end{center}
% \caption{The history of the $L_2$ norm of the model parameters. The model parameters $p$ is defined in Eq. \ref{eq:pdef} and the one for the trained model is presented in Table~\ref{tbl:p}}
% \label{figure:pnorm}
% \end{figure}

As far as the computational cost is concerned, SGD optimization is also much more efficient than genetic algorithms. To estimate the computational costs for both algorithms, we denote the computation of the IDT for all 96 samples as one computational unit. Since the computational cost for gradient evaluation is negligible, the current SGD optimization for F-24 costs about 100 units. For comparison, the GA optimization for the same datasets and base model costs about $10^5$ units, with 32 particles and 300 generations \cite{ryu2021data}. Such direct comparison reveals 1000 times of acceleration, which also results in a reduced need for computational power. For example, the current training was performed on a desktop workstation for hours, while the GA optimization in \cite{ryu2021data} was performed on a high-performance cluster. In addition, the acceleration factor scales with the size of the dataset and the dimension of the model parameter. Therefore, such computational saving is expected to be more significant when the kinetic system is more complex and more data are available.

%%%%%%%%%%%%%%%%%%%%%%%%%%%%%%%%%%%%%%%%%%%%%%%%%%%%%%%%%%%%%%%%%%%%%%
\section*{CONCLUSION}

This work presents a machine learning approach for learning and optimizing HyChem models. By treating the kinetic model as a neural network and using the stochastic gradient descent for optimization, HyChem models can be learned from ignition delay time measurements. Such an approach was demonstrated in training a HyChem model for the jet fuel of F-24. Compared to the commonly utilized genetic algorithms, the current approach has 1000 times less computational cost and better generalization performance with minimal degradation to the high-temperature chemistry while significantly improving the performance at low temperatures. In addition, further computational cost-saving is expected for larger datasets and more complex kinetic systems. Finally, the code for the current approach is built upon open-source packages and is available online.

While the current work focuses on developing HyChem models, such optimization techniques can be readily applied to other computational tasks. For example, optimizing overly reduced chemical models is a technique that has been pursued in practical engine simulations \cite{mittal2017multi,kelly2021toward}. In such tasks, high dimensional kinetic systems with hundreds of model parameters are usually present. Therefore, the current SGD-based approach will be superior to genetic algorithms.

%%%%%%%%%%%%%%%%%%%%%%%%%%%%%%%%%%%%%%%%%%%%%%%%%%%%%%%%%%%%%%%%%%%%%%
\begin{acknowledgment}

WJ and SD would like to acknowledge the funding support from Weichai Power Co., Ltd. WJ would like to acknowledge the help from Dr. Vyaas Gururajan on the implementation of the sensBVP method and the help from Dr. Je Ir Ryu on simulating F-24 ignition with the GA-optimized HyChem model.

\end{acknowledgment}

\bibliographystyle{asmems4}
\bibliography{ms}

\begin{thebibliography}{10}

\bibitem{violi2002experimental}
Violi, A., Yan, S., Eddings, E., Sarofim, A., Granata, S., Faravelli, T., and
  Ranzi, E., 2002.
\newblock ``Experimental formulation and kinetic model for jp-8 surrogate
  mixtures''.
\newblock {\em Combustion Science and Technology, \textbf{ 174}}(11-12),
  pp.~399--417.

\bibitem{eddings2005formulation}
Eddings, E.~G., Yan, S., Ciro, W., and Sarofim, A.~F., 2005.
\newblock ``Formulation of a surrogate for the simulation of jet fuel pool
  fires''.
\newblock {\em Combustion science and technology, \textbf{ 177}}(4),
  pp.~715--739.

\bibitem{dooley2010jet}
Dooley, S., Won, S.~H., Chaos, M., Heyne, J., Ju, Y., Dryer, F.~L., Kumar, K.,
  Sung, C.-J., Wang, H., Oehlschlaeger, M.~A., et~al., 2010.
\newblock ``A jet fuel surrogate formulated by real fuel properties''.
\newblock {\em Combustion and flame, \textbf{ 157}}(12), pp.~2333--2339.

\bibitem{wang2018physics}
Wang, H., Xu, R., Wang, K., Bowman, C.~T., Hanson, R.~K., Davidson, D.~F.,
  Brezinsky, K., and Egolfopoulos, F.~N., 2018.
\newblock ``A physics-based approach to modeling real-fuel combustion
  chemistry-i. evidence from experiments, and thermodynamic, chemical kinetic
  and statistical considerations''.
\newblock {\em Combustion and Flame, \textbf{ 193}}, pp.~502--519.

\bibitem{xu2018physics}
Xu, R., Wang, K., Banerjee, S., Shao, J., Parise, T., Zhu, Y., Wang, S.,
  Movaghar, A., Lee, D.~J., Zhao, R., et~al., 2018.
\newblock ``A physics-based approach to modeling real-fuel combustion
  chemistry--ii. reaction kinetic models of jet and rocket fuels''.
\newblock {\em Combustion and Flame, \textbf{ 193}}, pp.~520--537.

\bibitem{xu2020physics}
Xu, R., Saggese, C., Lawson, R., Movaghar, A., Parise, T., Shao, J., Choudhary,
  R., Park, J.-W., Lu, T., Hanson, R.~K., et~al., 2020.
\newblock ``A physics-based approach to modeling real-fuel combustion
  chemistry--vi. predictive kinetic models of gasoline fuels''.
\newblock {\em Combustion and Flame, \textbf{ 220}}, pp.~475--487.

\bibitem{ryu2021data}
Ryu, J.~I., Kim, K., Min, K., Scarcelli, R., Som, S., Kim, K.~S., Temme, J.~E.,
  Kweon, C.-B.~M., and Lee, T., 2021.
\newblock ``Data-driven chemical kinetic reaction mechanism for f-24 jet fuel
  ignition''.
\newblock {\em Fuel, \textbf{ 290}}, p.~119508.

\bibitem{ji2021autonomous}
Ji, W., and Deng, S., 2021.
\newblock ``Autonomous discovery of unknown reaction pathways from data by
  chemical reaction neural network''.
\newblock {\em The Journal of Physical Chemistry A, \textbf{ 125}}(4),
  pp.~1082--1092.

\bibitem{chen2018neural}
Chen, R.~T., Rubanova, Y., Bettencourt, J., and Duvenaud, D., 2018.
\newblock ``Neural ordinary differential equations''.
\newblock {\em arXiv preprint arXiv:1806.07366}.

\bibitem{baydin2018automatic}
Baydin, A.~G., Pearlmutter, B.~A., Radul, A.~A., and Siskind, J.~M., 2018.
\newblock ``Automatic differentiation in machine learning: a survey''.
\newblock {\em Journal of machine learning research, \textbf{ 18}}.

\bibitem{abadi2016tensorflow}
Abadi, M., Barham, P., Chen, J., Chen, Z., Davis, A., Dean, J., Devin, M.,
  Ghemawat, S., Irving, G., Isard, M., et~al., 2016.
\newblock ``Tensorflow: A system for large-scale machine learning''.
\newblock In 12th $\{$USENIX$\}$ symposium on operating systems design and
  implementation ($\{$OSDI$\}$ 16), pp.~265--283.

\bibitem{jax2018github}
Bradbury, J., Frostig, R., Hawkins, P., Johnson, M.~J., Leary, C., Maclaurin,
  D., Necula, G., Paszke, A., Vander{P}las, J., Wanderman-{M}ilne, S., and
  Zhang, Q., 2018.
\newblock {JAX}: composable transformations of {P}ython+{N}um{P}y programs.

\bibitem{paszke2019pytorch}
Paszke, A., Gross, S., Massa, F., Lerer, A., Bradbury, J., Chanan, G., Killeen,
  T., Lin, Z., Gimelshein, N., Antiga, L., et~al., 2019.
\newblock ``Pytorch: An imperative style, high-performance deep learning
  library''.
\newblock {\em arXiv preprint arXiv:1912.01703}.

\bibitem{revels2016forward}
Revels, J., Lubin, M., and Papamarkou, T., 2016.
\newblock ``Forward-mode automatic differentiation in julia''.
\newblock {\em arXiv preprint arXiv:1607.07892}.

\bibitem{innes2018don}
Innes, M., 2018.
\newblock ``Don't unroll adjoint: Differentiating ssa-form programs''.
\newblock {\em arXiv preprint arXiv:1810.07951}.

\bibitem{jiarrheniusgithub}
Ji, W., and Deng, S., 2021.
\newblock Arrhenius.jl: A differentiable combustion simulation package.
\newblock \url{https://github.com/DENG-MIT/Arrhenius.jl}.

\bibitem{ji2019evolution}
Ji, W., Ren, Z., and Law, C.~K., 2019.
\newblock ``Evolution of sensitivity directions during autoignition''.
\newblock {\em Proceedings of the Combustion Institute, \textbf{ 37}}(1),
  pp.~807--815.

\bibitem{kee1989chemkin}
Kee, R.~J., Rupley, F.~M., and Miller, J.~A., 1989.
\newblock Chemkin-ii: A fortran chemical kinetics package for the analysis of
  gas-phase chemical kinetics.
\newblock Tech. rep., Sandia National Lab.(SNL-CA), Livermore, CA (United
  States).

\bibitem{goodwin2009cantera}
Goodwin, D.~G., Moffat, H.~K., and Speth, R.~L., 2009.
\newblock Cantera: An object-oriented software toolkit for chemical kinetics,
  thermodynamics, and transport processes.

\bibitem{rackauckas2018comparison}
Rackauckas, C., Ma, Y., Dixit, V., Guo, X., Innes, M., Revels, J., Nyberg, J.,
  and Ivaturi, V., 2018.
\newblock ``A comparison of automatic differentiation and continuous
  sensitivity analysis for derivatives of differential equation solutions''.
\newblock {\em arXiv preprint arXiv:1812.01892}.

\bibitem{rackauckas2017differentialequations}
Rackauckas, C., and Nie, Q., 2017.
\newblock ``Differentialequations. jl--a performant and feature-rich ecosystem
  for solving differential equations in julia''.
\newblock {\em Journal of Open Research Software, \textbf{ 5}}(1).

\bibitem{rackauckas2020universal}
Rackauckas, C., Ma, Y., Martensen, J., Warner, C., Zubov, K., Supekar, R.,
  Skinner, D., Ramadhan, A., and Edelman, A., 2020.
\newblock ``Universal differential equations for scientific machine learning''.
\newblock {\em arXiv preprint arXiv:2001.04385}.

\bibitem{raissi2019physics}
Raissi, M., Perdikaris, P., and Karniadakis, G.~E., 2019.
\newblock ``Physics-informed neural networks: A deep learning framework for
  solving forward and inverse problems involving nonlinear partial differential
  equations''.
\newblock {\em Journal of Computational Physics, \textbf{ 378}}, pp.~686--707.

\bibitem{ji2020stiff}
Ji, W., Qiu, W., Shi, Z., Pan, S., and Deng, S., 2020.
\newblock ``Stiff-pinn: Physics-informed neural network for stiff chemical
  kinetics''.
\newblock {\em arXiv preprint arXiv:2011.04520}.

\bibitem{allen2012application}
Allen, C., Toulson, E., Edwards, T., and Lee, T., 2012.
\newblock ``Application of a novel charge preparation approach to testing the
  autoignition characteristics of jp-8 and camelina hydroprocessed renewable
  jet fuel in a rapid compression machine''.
\newblock {\em Combustion and Flame, \textbf{ 159}}(9), pp.~2780--2788.

\bibitem{gururajan2019direct}
Gururajan, V., and Egolfopoulos, F.~N., 2019.
\newblock ``Direct sensitivity analysis for ignition delay times''.
\newblock {\em Combustion and Flame, \textbf{ 209}}, pp.~478--480.

\bibitem{lemke2019adjoint}
Lemke, M., Cai, L., Reiss, J., Pitsch, H., and Sesterhenn, J., 2019.
\newblock ``Adjoint-based sensitivity analysis of quantities of interest of
  complex combustion models''.
\newblock {\em Combustion Theory and Modelling, \textbf{ 23}}(1), pp.~180--196.

\bibitem{almohammadi2021tangent}
Almohammadi, S., Hantouche, M., Le~Ma{\^\i}tre, O.~P., and Knio, O.~M., 2021.
\newblock ``A tangent linear approximation of the ignition delay time. i:
  Sensitivity to rate parameters''.
\newblock {\em Combustion and Flame, \textbf{ 230}}, p.~111426.

\bibitem{kingma2014adam}
Kingma, D.~P., and Ba, J., 2014.
\newblock ``Adam: A method for stochastic optimization''.
\newblock {\em arXiv preprint arXiv:1412.6980}.

\bibitem{bengio2017deep}
Bengio, Y., Goodfellow, I., and Courville, A., 2017.
\newblock {\em Deep learning}, Vol.~1.
\newblock MIT press Massachusetts, USA:.

\bibitem{ji2017crossover}
Ji, W., Zhao, P., Zhang, P., Ren, Z., He, X., and Law, C.~K., 2017.
\newblock ``On the crossover temperature and lower turnover state in the ntc
  regime''.
\newblock {\em Proceedings of the Combustion Institute, \textbf{ 36}}(1),
  pp.~343--353.

\bibitem{zhao2013role}
Zhao, P., and Law, C.~K., 2013.
\newblock ``The role of global and detailed kinetics in the first-stage
  ignition delay in ntc-affected phenomena''.
\newblock {\em Combustion and Flame, \textbf{ 160}}(11), pp.~2352--2358.

\bibitem{zhang2016first}
Zhang, P., Ji, W., He, T., He, X., Wang, Z., Yang, B., and Law, C.~K., 2016.
\newblock ``First-stage ignition delay in the negative temperature coefficient
  behavior: Experiment and simulation''.
\newblock {\em Combustion and Flame, \textbf{ 167}}, pp.~14--23.

\bibitem{he2005experimental}
He, X., Donovan, M., Zigler, B., Palmer, T., Walton, S., Wooldridge, M., and
  Atreya, A., 2005.
\newblock ``An experimental and modeling study of iso-octane ignition delay
  times under homogeneous charge compression ignition conditions''.
\newblock {\em Combustion and Flame, \textbf{ 142}}(3), pp.~266--275.

\bibitem{ji2015intermediate}
Ji, W., Zhang, P., He, T., Wang, Z., Tao, L., He, X., and Law, C.~K., 2015.
\newblock ``Intermediate species measurement during iso-butanol
  auto-ignition''.
\newblock {\em Combustion and Flame, \textbf{ 162}}(10), pp.~3541--3553.

\bibitem{gao2019global}
Gao, X., Gou, X., and Sun, W., 2019.
\newblock ``Global pathway analysis: a hierarchical framework to understand
  complex chemical kinetics''.
\newblock {\em Combustion Theory and Modelling, \textbf{ 23}}(3), pp.~549--571.

\bibitem{long2019numerical}
Long, A.~E., Speth, R.~L., and Green, W.~H., 2019.
\newblock ``Numerical investigation of strained extinction at engine-relevant
  pressures: Pressure dependence and sensitivity to chemical and physical
  parameters for methane-based flames''.
\newblock {\em Combustion and Flame, \textbf{ 202}}, pp.~318--333.

\bibitem{ji2020dependence}
Ji, W., Yang, T., Ren, Z., and Deng, S., 2020.
\newblock ``Dependence of kinetic sensitivity direction in premixed flames''.
\newblock {\em Combustion and Flame, \textbf{ 220}}, pp.~16--22.

\bibitem{mittal2017multi}
Mittal, A., Wijeyakulasuriya, S.~D., Probst, D., Banerjee, S., Finney, C.~E.,
  Edwards, K.~D., Willcox, M., and Naber, C., 2017.
\newblock ``Multi-dimensional computational combustion of highly dilute,
  premixed spark-ignited opposed-piston gasoline engine using direct chemistry
  with a new primary reference fuel mechanism''.
\newblock In Internal Combustion Engine Division Fall Technical Conference,
  Vol.~58325, American Society of Mechanical Engineers, p.~V002T06A022.

\bibitem{kelly2021toward}
Kelly, M., Dooley, S., and Bourque, G., 2021.
\newblock ``Toward machine learned highly reduce kinetic models for methane/air
  combustion''.
\newblock {\em arXiv preprint arXiv:2103.08377}.

\end{thebibliography}

\end{document}